\documentclass[journal,a4paper]{IEEEtran}

\usepackage[english]{babel}
\usepackage[T1]{fontenc}
\usepackage{newtxtext,newtxmath} 
\usepackage{subfig}

\usepackage{graphicx}
\usepackage{booktabs,array}

\usepackage{siunitx}
\usepackage{cite}
\usepackage{url}

\usepackage{algorithm}
\usepackage{algpseudocode}
\usepackage[font=small,labelfont=bf]{caption}
\usepackage{subcaption} 

\usepackage{xcolor}
\usepackage[most]{tcolorbox}

\definecolor{promptgray}{gray}{0.94}
\definecolor{promptgreen}{RGB}{0,100,0}
\definecolor{promptblue}{RGB}{0,60,140}
\definecolor{promptorange}{RGB}{150,70,0}

\tcbset{
  colback=promptgray,
  colframe=black!60,
  boxrule=0.3pt,
  arc=2pt,
  left=4pt, right=4pt, top=4pt, bottom=4pt,
}

\newtcolorbox{llmprompt}[2][]{title={#2},#1}
\newcommand{\query}[1]{\textcolor{promptgreen}{\texttt{#1}}}
\newcommand{\resp}[1]{\textcolor{promptblue}{\texttt{#1}}}

\usepackage{amsthm}
\theoremstyle{definition}

\newtheorem{assumption}{Assumption}

\newcommand{\be}{\begin{equation}}
\newcommand{\ee}{\end{equation}}

\begin{document}

 \title{Generalized Intelligence for Tactical Decision-Making: Large Language Model–Driven Dynamic Weapon Target Assignment}




\author{Johannes Autenrieb$^{1}$, Ole Ostermann$^{1}$
\thanks{$^{1}$ German Aerospace Center (DLR), Institute of Flight Systems, 38108, Braunschweig, Germany.
(email: \texttt{johannes.autenrieb@dlr.de, ole.ostermann@dlr.de)}
}}




\maketitle

\begin{abstract}
Modern aerospace defense systems increasingly rely on autonomous decision-making to coordinate large numbers of interceptors against multiple incoming threats. Conventional weapon–target assignment (WTA) algorithms, including mixed-integer programming and auction-based methods, show limitations in dynamic and uncertain tactical environments where human-like reasoning and adaptive prioritization are required. This paper introduces a large language model (LLM) driven WTA framework that integrates generalized intelligence into cooperative missile guidance. The proposed system formulates the tactical decision process as a reasoning problem, in which an LLM evaluates spatial and temporal relationships among interceptors, targets, and defended assets to generate real-time assignments. In contrast to classical optimization methods, the approach leverages contextual mission data such as threat direction, asset priority, and closing velocity to adapt dynamically and reduce assignment switching. A dedicated simulation environment supports both static and dynamic assignment modes. Results demonstrate improved consistency, adaptability, and mission-level prioritization, establishing a foundation for integrating generalized artificial intelligence into tactical guidance systems.
\end{abstract}


\section{Introduction}
Modern military engagements have grown increasingly complex with advanced missile technologies, swarm tactics, and deceptive strategies, making the efficient allocation of defensive assets a critical challenge in aerospace operations~\cite{Li2024}. 
The weapon target assignment (WTA) problem seeks to allocate a set of interceptors to a set of incoming threats under operational, geometric, and temporal constraints to maximize mission effectiveness. 
Originally formulated as a combinatorial optimization problem~\cite{manne1958wta}, WTA has been recognized as NP-complete~\cite{Lloyd1986}, leading to a long history of research into exact, heuristic, and meta-heuristic algorithms for both static and dynamic scenarios~\cite{Li2024,bertsekas1988auction,pentico2007assignment}. 
In realistic mission environments, the problem becomes even more challenging due to coupling between assignment and guidance, uncertainty in target motion, and time-critical decision requirements.

Classical formulations commonly decouple the assignment and trajectory optimization stages, solving them sequentially to maintain tractability. 
This simplification, however, often limits optimality and situational adaptability. 
Recent work introduced integrated optimization frameworks that unify both target assignment and trajectory planning within a single decision process~\cite{Jin2025}. 
By embedding allocation decisions within a continuous optimization routine, these methods jointly account for geometry, timing, and dynamic feasibility, demonstrating the advantages of coordinated reasoning across hierarchical layers of the engagement problem.

Parallel to advances in optimization, recent studies have explored the use of data-driven and machine learning (ML) techniques to improve adaptability in WTA~\cite{Li2024}. 
Reinforcement learning and graph-based architectures have shown promise for dynamic allocation, yet their reliance on large-scale training data and limited interpretability restricts their deployment in safety-critical missions~\cite{shokoohi2022rl}. 
These limitations motivate the exploration of hybrid reasoning systems that combine algorithmic precision with contextual understanding and explainability.

\begin{figure}
    \centering
    \includegraphics[width=\columnwidth]{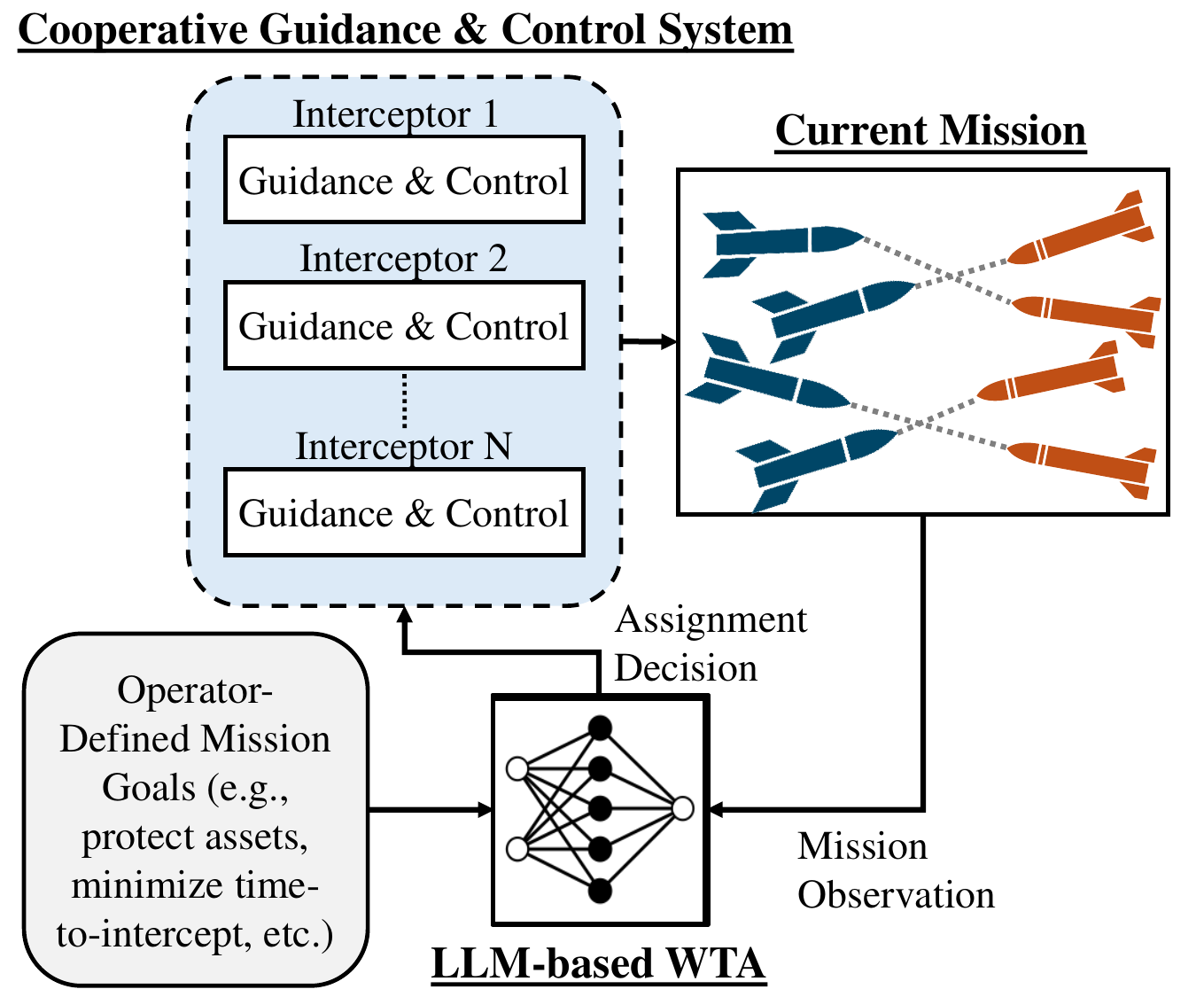}
    \caption{Illustration of the proposed LLM–Driven WTA architecture.}
    \label{fig:overview_concept}
\end{figure}

Large language models (LLMs) have recently emerged as a new class of reasoning systems with potential applications in autonomous decision-making. 
Pre-trained on large-scale multimodal data, LLMs can process numerical and symbolic inputs jointly, enabling them to perform high-level reasoning and task decomposition without explicit cost function definitions~\cite{Pallagani2024}. 
Their integration into control and planning pipelines has opened new avenues for mission-level decision support, particularly where symbolic knowledge and numerical optimization intersect. 
Notably, recent research demonstrated that LLMs can be embedded directly into robotic feedback loops to enhance resilience and adaptation~\cite{Tagliabue2023}. 
In that work, the LLM contributed to mission planning, state interpretation, and control adjustment, reducing errors and preventing unsafe behaviors even under unmodeled dynamics. 
This evidence suggests that LLMs can complement algorithmic decision-making by providing contextual reasoning where traditional models are limited.

Motivated by these developments, this work investigates the use of LLM-based reasoning for dynamic weapon–target assignment and cooperative missile guidance. 
The proposed framework treats assignment generation as a contextual reasoning task, where the LLM interprets the global mission state—including geometry, timing, and priority metrics—and outputs feasible interceptor–target allocations without predefined weighting parameters. 
By integrating LLM-based decision support into the assignment–guidance loop, the approach preserves the structure of classical guidance laws while leveraging the LLM’s ability to reason about dynamic mission context. 
The resulting system aims to bridge the gap between numerical optimization and human-level reasoning, offering interpretable, adaptive, and context-aware solutions for multi-interceptor coordination under uncertainty.

\section{Problem Formulation}
\label{sec:problem}
We consider a set of interceptors $\mathcal{M}=\{1,\dots,N\}$ and a set of targets $\mathcal{T}=\{1,\dots,N_T\}$ evolving in three-dimensional Euclidean space $\mathbb{R}^3$. 
Each interceptor $i\in\mathcal{M}$ is modeled as a nonlinear control-affine system of the form
\begin{equation}
\dot{\mathbf{x}}_i(t) = \mathbf{f}\big(\mathbf{x}_i(t)\big) + \mathbf{g}\big(\mathbf{x}_i(t)\big)\mathbf{u}_i(t),
\label{eq:interceptor_dyn}
\end{equation}
where $\mathbf{x}_i\in\mathbb{R}^{n_x}$ denotes the interceptor state vector, and $\mathbf{u}_i\in\mathbb{R}^{n_u}$ the commanded control input. 
The functions $\mathbf{f}(\cdot)$ and $\mathbf{g}(\cdot)$ represent the nonlinear drift and control input dynamics, respectively, and are assumed to be locally Lipschitz continuous to guarantee existence and uniqueness of trajectories.  
This formulation encompasses a broad class of interceptor dynamics, including translational motion with aerodynamic and thrust-vector effects, while retaining control-affine structure for analytical and numerical tractability.

Targets $k\in\mathcal{T}$ are modeled analogously as
\begin{equation}
\dot{\mathbf{x}}^t_k(t)=\mathbf{f}^t\big(\mathbf{x}^t_k(t)\big)+\mathbf{g}^t\big(\mathbf{x}^t_k(t)\big)\mathbf{u}^t_k(t),
\label{eq:target_dyn}
\end{equation}
where $\mathbf{u}^t_k(t)$ represents a known or bounded target maneuver input.  
All states and control inputs are subject to compact physical bounds,
\begin{equation}
\|\mathbf{u}_i(t)\|\le a_{\max}, 
\qquad 
\|\mathbf{x}_i(t)\|\le x_{\max},
\label{eq:constraints}
\end{equation}
for all $t\in[t_0,t_f]$.  
The nonlinear control-affine representation~\eqref{eq:interceptor_dyn}–\eqref{eq:target_dyn} provides a general and flexible basis for describing interceptor–target dynamics, encompassing the linear double-integrator model as a special case when $\mathbf{f}(\mathbf{x}) = \mathbf{A}\mathbf{x}$ and $\mathbf{g}(\mathbf{x}) = \mathbf{B}$.

The primary objective of the mission is that each interceptor reaches an assigned target. 
The nominal guidance law is based on \emph{Proportional Navigation Guidance} (PNG), which serves as a benchmark strategy minimizing the line-of-sight (LOS) rotation rate and achieving near-optimal interception for non-maneuvering or mildly maneuvering targets.

For each interceptor–target pair $(i,k)$, the relative position and velocity vectors are defined as
\begin{equation}
\mathbf{r}_{ik}(t)=\mathbf{p}_i(t)-\mathbf{p}^t_k(t),
\qquad
\mathbf{v}_{ik}(t)=\mathbf{v}_i(t)-\mathbf{v}^t_k(t),
\label{eq:relative_def}
\end{equation}
and the unit line-of-sight vector as
\begin{equation}
\hat{\mathbf{r}}_{ik}(t)=\frac{\mathbf{r}_{ik}(t)}{\|\mathbf{r}_{ik}(t)\|}.
\end{equation}
The LOS angular-rate vector is given by
\begin{equation}
\dot{\boldsymbol{\lambda}}_{ik}(t)=
\frac{\mathbf{r}_{ik}(t)\times \mathbf{v}_{ik}(t)}{\|\mathbf{r}_{ik}(t)\|^2}.
\label{eq:los_rate}
\end{equation}
The PNG acceleration command then follows as
\begin{equation}
\mathbf{u}_i(t)=
N\,\|\mathbf{v}_{ik}(t)\|\,\big(\dot{\boldsymbol{\lambda}}_{ik}(t)\times \hat{\mathbf{r}}_{ik}(t)\big),
\label{eq:png}
\end{equation}
where $N>0$ is the navigation constant. 
The commanded acceleration lies orthogonal to the LOS and drives the LOS rate toward zero, thereby aligning the interceptor’s velocity vector with the LOS. 
Equation~\eqref{eq:png} is subsequently saturated according to~\eqref{eq:constraints} to ensure feasibility.

\medskip
The overall engagement problem can be formulated as a finite-horizon mixed-integer optimal control problem. 
Let binary variables $z_{ik}\in\{0,1\}$ indicate whether interceptor~$i$ is assigned to target~$k$. 
The assignment matrix $\mathbf{Z}=[z_{ik}]$ satisfies
\begin{equation}
\sum_{k=1}^{N_T} z_{ik}=1, 
\qquad 
\sum_{i=1}^{N} z_{ik}\ge 1,
\label{eq:assign_constraints}
\end{equation}
ensuring that each interceptor is assigned to exactly one target, while each target ideally receives at least one interceptor.  
The second condition is enforced only when sufficient resources are available ($N \ge N_T$); otherwise, it can be relaxed or adapted prior to optimization to reflect resource-limited scenarios.  
This adaptive constraint setup enables asymmetric mission configurations, where redundant interceptors can be cooperatively allocated to a single target to increase interception probability and robustness.

Over the mission horizon $\mathcal{H}=[t_0,t_f]$, the continuous-time mixed-integer optimal control problem (MIOCP) is defined as
\begin{equation}
\begin{aligned}
\min_{\{\mathbf{u}_i(\cdot)\},\,\mathbf{Z}} \quad
& J = 
\sum_{i=1}^{N}\sum_{k=1}^{N_T} 
\int_{t_0}^{t_f}
c_{ik}\!\big(\mathbf{x}_i(t),\mathbf{x}^t_k(t),\mathbf{u}_i(t),z_{ik}\big)\,dt \\[3pt]
\text{subject to}\quad
& \dot{\mathbf{x}}_i(t)=\mathbf{f}\big(\mathbf{x}_i(t)\big)+\mathbf{g}\big(\mathbf{x}_i(t)\big)\mathbf{u}_i(t),\\[3pt]
& \dot{\mathbf{x}}^t_k(t)=\mathbf{f}^t\big(\mathbf{x}^t_k(t)\big)+\mathbf{g}^t\big(\mathbf{x}^t_k(t)\big)\mathbf{u}^t_k(t),\\[3pt]
&\mathbf{x}_i(t_0)=\mathbf{x}_{i,0},\quad
\mathbf{x}^t_k(t_0)=\mathbf{x}^t_{k,0}, \\[3pt]
& \|\mathbf{u}_i(t)\|\le a_{\max},\quad
\|\mathbf{x}_i(t)\|\le x_{\max},\forall\,t\in[t_0,t_f],\\[3pt]
& z_{ik}\in\{0,1\},\quad\text{satisfying \eqref{eq:assign_constraints}}.
\end{aligned}
\label{eq:ocp}
\end{equation}

Here, $c_{ik}(\cdot)$ denotes the mission cost function associated with each interceptor–target pair, which may represent, for example, a multi-metric formulation combining mission goals and resource considerations.  
The overall mission cost $J$ thus captures the coupled interaction between assignment and guidance decisions across all interceptors and targets.  
Through the minimization in~\eqref{eq:ocp}, the objective is to reduce the global mission cost by optimizing both the discrete assignment $\mathbf{Z}$ and the continuous control inputs $\mathbf{u}_i(\cdot)$.

However, the resulting optimization problem~\eqref{eq:ocp} is NP-hard, as it combines continuous and discrete decision spaces under nonlinear coupling.  
Even for moderate system sizes, direct real-time solution is computationally intractable.  
Therefore, practical implementations often employ surrogate or decomposed formulations that separate the assignment and guidance components while retaining near-optimal performance.  
These approaches, including linear assignment, MILP-based, and auction-based algorithms, are discussed in the following section.

\section{Comparative Overview of Weapon--Target Assignment Methods}
\label{sec:wta}
As established by the continuous-time MIOCP in~\eqref{eq:ocp}, jointly optimizing assignment and guidance is nonlinear, nonconvex, and NP-hard, which prevents real-time solution at realistic scales. Practical implementations therefore adopt surrogate models that separate the discrete assignment from the continuous guidance and evaluate pairwise interceptor--target scores built from instantaneous kinematic quantities and mission parameters.

\begin{assumption}[Pairwise Separable Cost Surrogates]
\label{ass:pairwise}
The global mission cost can be approximated as the sum of independent pairwise costs,
\begin{equation}
J \approx 
\sum_{i=1}^{N}\sum_{k=1}^{N_T} c_{ik}\, z_{ik},
\label{eq:pairwise-sum}
\end{equation}
where each $c_{ik}$ depends only on the kinematic state of interceptor~$i$ and target~$k$ and on local mission weights, but is independent of other assignments. 
This assumption neglects interdependencies such as mutual collision avoidance, fuel coupling, or cooperative interception geometry among simultaneous engagements. 
It is, however, the foundation of most tractable WTA formulations in operations research and control, enabling a linear surrogate representation \cite{DLR_Autenrieb2019}.
\end{assumption}

\begin{assumption}[Assignment--Guidance Decomposition]
\label{ass:decomposition}
The binary assignment variables $\{z_{ik}\}$ are held constant over the mission horizon $\mathcal{T}$ (static assignment) or updated only at discrete epochs $t_h = t_0 + h\,\Delta T$ (dynamic reassignment). 
Between updates, each active pair $(i,k)$ evolves according to the continuous dynamics~\eqref{eq:interceptor_dyn} and the nominal guidance law~\eqref{eq:png}, subject to the bounds in~\eqref{eq:constraints}.
\end{assumption}

Under Assumptions~\ref{ass:pairwise}--\ref{ass:decomposition}, the real mission cost can be approximated by a weighted combination of geometric, kinematic, and mission-level metrics. For each interceptor--target pair $(i,k)$, we therefore model an instantaneous surrogate cost that captures the most relevant factors for local engagement decisions. Accordingly, we approximate the per-pair cost as
\begin{equation}
\label{eq:pairwise-cost}
c_{ik} =
w_d\,\|\mathbf{r}_{ik}\|
+w_v\,\|\mathbf{v}_{ik}\|
+w_\theta\,\theta_k
+w_\psi\,\psi_k,
\end{equation}
with
\begin{equation*}
w_d,w_v,w_\theta,w_\psi>0,
\end{equation*}
where $\|\mathbf{r}_{ik}\|$ and $\|\mathbf{v}_{ik}\|$ denote the Euclidean norms of the relative position and velocity vectors defined in~\eqref{eq:relative_def}. The terms $\theta_k$ and $\psi_k$ are target-dependent importance metrics: $\theta_k$ represents the intrinsic priority or threat level of target~$k$, while $\psi_k$ encodes contextual mission relevance, such as proximity to defended assets or predicted time-to-impact. The weighting factors $(w_d,w_v,w_\theta,w_\psi)$ determine the relative contribution of each metric within the surrogate formulation.

Stacking all pairwise costs yields the cost matrix $\mathbf{C} = [c_{ik}] \in \mathbb{R}^{N \times N_T}$, which aggregates the instantaneous engagement costs between interceptors $i \in \mathcal{M}$ and targets $k \in \mathcal{T}$. The resulting global assignment problem is formulated as a linear assignment program (LAP):
\begin{equation}
\label{eq:lap}
\begin{aligned}
\min_{\{z_{ik}\}} \quad
& \sum_{i=1}^{N}\sum_{k=1}^{N_T} c_{ik}\, z_{ik} \\[4pt]
\text{subject to} \quad
& \sum_{k=1}^{N_T} z_{ik} = 1, 
\quad i = 1,\dots,N,\\[3pt]
& z_{ik} \in \{0,1\}.
\end{aligned}
\end{equation}
Each binary variable $z_{ik}$ indicates whether interceptor~$i$ is assigned to target~$k$. The equality constraint ensures that every interceptor is assigned to exactly one target, while the formulation allows several interceptors to engage the same target when this minimizes the overall mission cost. This structure enables asymmetric defense scenarios where the number of interceptors exceeds the number of incoming threats, allowing the system to exploit numerical superiority or redundancy to increase interception probability and selectively disregard low-priority decoys  \cite{DLR_Ostermann2025}. It also supports cooperative engagement strategies in which multiple interceptors contribute to the protection of high-value assets.

\subsection*{A. Linear Assignment via the Hungarian Algorithm}
The Hungarian method~\cite{kuhn1955hungarian,munkres1957alg} provides a deterministic global optimum for the special case of the linear assignment program~\eqref{eq:lap} where the cost matrix $\mathbf{C} = [c_{ik}] \in \mathbb{R}^{N \times N_T}$ is square, that is, when the number of interceptors and targets is equal. With a computational complexity of $\mathcal{O}(n^3)$, $n=\max\{N,N_T\}$, the method achieves excellent numerical stability and remains well suited for centralized and reproducible engagement scenarios.

A main limitation of the Hungarian algorithm is that it directly solves the global optimization problem. Consequently, all interceptors must have access to the full cost matrix $\mathbf{C}$, requiring centralized information sharing or a consensus mechanism to ensure consistent global decisions. In practice, this means that either a central coordinator computes and broadcasts the assignment to all interceptors, or each interceptor independently solves the same problem using synchronized mission data.

As discussed, the classical Hungarian algorithm assumes a square cost matrix. However, several extensions and modern variants~\cite{pentico2007assignment} have been developed to handle rectangular or asymmetric cases, enabling its application to scenarios with differing numbers of interceptors and targets.

\subsection*{B. Mixed–Integer Linear Programming (MILP)}
The mixed–integer linear program (MILP) extends the linear assignment formulation~\eqref{eq:lap} by allowing additional continuous and integer decision variables to model coupled mission-level logic. In contrast to the Hungarian algorithm, the MILP can handle rectangular cost matrices $\mathbf{C} = [c_{ik}] \in \mathbb{R}^{N \times N_T}$ directly and imposes no restriction on the relative number of interceptors and targets. The general problem can be expressed as
\begin{equation}
\label{eq:milp}
\begin{aligned}
\min_{\{z_{ik}\},\,\mathbf{y}} \quad
& \sum_{i=1}^{N}\sum_{k=1}^{N_T} c_{ik}\, z_{ik} + \mathbf{f}^\top \mathbf{y} \\[2pt]
\text{subject to}\quad
& \mathbf{A}_{\mathrm{eq}}
\begin{bmatrix}
\mathrm{vec}(\mathbf{Z})\\[2pt]
\mathbf{y}
\end{bmatrix}
=
\mathbf{b}_{\mathrm{eq}},\,\,
\mathbf{A}_{\mathrm{ineq}}
\begin{bmatrix}
\mathrm{vec}(\mathbf{Z})\\[2pt]
\mathbf{y}
\end{bmatrix}
\le
\mathbf{b}_{\mathrm{ineq}},\\[3pt]
& z_{ik} \in \{0,1\}, \quad \mathbf{y} \ge 0,
\end{aligned}
\end{equation}
where $\mathbf{z} = \mathrm{vec}(\mathbf{Z})$ represents the stacked binary assignment variables and $\mathbf{y}$ collects continuous auxiliary variables associated with mission-level quantities such as time-to-intercept, range margins, or engagement overlap. The vector $\mathbf{f}$ defines their corresponding cost coefficients.

The matrices $\mathbf{A}_{\mathrm{eq}}$ and $\mathbf{A}_{\mathrm{ineq}}$, together with the vectors $\mathbf{b}_{\mathrm{eq}}$ and $\mathbf{b}_{\mathrm{ineq}}$, encode the specific mission constraints to be satisfied. Their structure depends on the considered engagement scenario and may include, for example, launcher or interceptor capacities, range and timing requirements, or compatibility restrictions between interceptor–target pairs. These matrices thus extend the earlier linear assignment problem~\eqref{eq:lap} to capture more complex mission-level constraints that cannot be represented by the cost function alone.

The MILP formulation preserves global optimality for the linear surrogate cost~\eqref{eq:pairwise-cost} while enabling considerably greater modeling flexibility. Unlike the Hungarian algorithm, which can only approximate such effects through cost weighting, the MILP enforces them as hard constraints or coupled continuous variables within the optimization process. Although MILPs are typically solved in a centralized manner using branch-and-bound or cutting-plane algorithms and their computational cost grows rapidly with problem size, they provide a general and expressive framework for realistic weapon–target assignment and related dynamic optimization problems~\cite{manne1958wta,hosein1990dynamic}.

\subsection*{C. Auction–Based Distributed Methods}
Auction algorithms, such as those introduced in~\cite{bertsekas1988auction,BRAQUET2021}, provide a decentralized, iterative approximation of the linear assignment problem~\eqref{eq:lap}.  
Instead of solving the optimization problem centrally, each interceptor locally evaluates its potential target allocations and exchanges information with neighboring agents or a limited communication network.  
Through iterative bidding and adjustment of local assignment preferences, the group collectively converges toward a consistent assignment without requiring access to the full global cost matrix.

In contrast to the Hungarian and MILP approaches, which depend on centralized information and coordination, auction-based methods achieve scalability and robustness through distributed consensus.  
They are particularly suited for large-scale or swarm defense applications, where communication may be intermittent and full network synchronization cannot be guaranteed.  
The decentralized structure makes them computationally efficient and fault-tolerant, but at the cost of losing deterministic optimality.  
Their convergence properties depend on the communication topology and tuning parameters, and the resulting assignments represent high-quality approximations rather than guaranteed global optima.  
Despite these limitations, auction-based approaches remain an effective strategy for achieving scalable and resilient coordination in cooperative engagement scenarios.

\medskip
In summary, the Hungarian algorithm~\cite{kuhn1955hungarian,munkres1957alg} and MILP formulations~\cite{manne1958wta,hosein1990dynamic} constitute centralized, deterministic baselines that optimize the separable surrogate implied by Assumption~\ref{ass:pairwise}, whereas auction methods~\cite{bertsekas1988auction} enable scalable, communication-limited consensus at the cost of weaker optimality guarantees.  
Across all approaches, however, several limitations persist: the independence assumption in~\eqref{eq:pairwise-sum} neglects geometric and temporal coupling between simultaneous engagements; linear surrogate costs can mis-rank feasible assignments when nonlinear dynamics dominate; and dynamic reassignment may exhibit oscillatory tasking without context-aware reasoning.  
A paradigm that relaxes Assumption~\ref{ass:pairwise}---while maintaining the temporal separation of Assumption~\ref{ass:decomposition}---could thus capture the coupled structure of multi-agent engagements more faithfully.  
In the following section, we investigate whether large language models, trained on broad human reasoning data, can provide such a context-informed assignment policy that complements traditional optimization-based solvers.

\section{LLM-Driven Dynamic Weapon--Target Assignment Framework}
\label{sec:llm_method}
\definecolor{promptgreen}{RGB}{0,100,0}
\definecolor{promptblue}{RGB}{0,60,140}

As discussed in the previous sections, the continuous-time MIOCP in~\eqref{eq:ocp} is intractable for real-time scales. Classical solutions (Hungarian, MILP, and auction; cf.\ Section~\ref{sec:wta}) therefore adopt the pairwise separability in Assumption~\ref{ass:pairwise} and solve variants of~\eqref{eq:lap} (or~\eqref{eq:milp}) after selecting weights for the surrogate in~\eqref{eq:pairwise-cost}. This enables fast computation but forces a fixed trade-off across geometry, timing, and priority, and it neglects interdependencies among simultaneous engagements. The proposed approach replaces the explicit surrogate by a reasoning-based assignment policy implemented with a large language model, while retaining the assignment–guidance decomposition in Assumption~\ref{ass:decomposition} so that continuous execution still follows PNG~\eqref{eq:png} between decision epochs. In other words, Assumption~\ref{ass:pairwise} is relaxed while the temporal separation is preserved.

We consider in this work a defended-asset scenario under the same kinematics and bounds as in Section~\ref{sec:problem}, with all positions and velocities expressed in a shared global Cartesian frame over the defended area (e.g., local NED). A centralized coordinator is assumed to have full situational awareness of interceptor, target, and asset states. The goal is to protect high-priority assets, reduce time-to-intercept, and avoid frequent reassignments. In a classical pipeline, one would instantiate~\eqref{eq:pairwise-cost} for each $(i,k)$ and tune $(w_d,w_v,w_\theta,w_\psi)$ to capture the desired trade-offs. These parameters, however, are brittle and scenario-dependent. Here, the same quantitative information is provided to the LLM, which reasons over the mission context to produce a feasible assignment vector without any predefined weights.

At discrete decision times $t_h$, the simulation aggregates all relevant scene data, constructs a structured prompt $\mathcal{P}_h$, and queries the LLM via an HTTP/JSON interface. The returned assignment vector is parsed, validated, and applied until the next update, during which PNG~\eqref{eq:png} governs the motion of each assigned interceptor–target pair. 

Regular expressions (RegEx) are extensively employed for string manipulation, extraction, and format validation to ensure that only syntactically correct MATLAB-style row vectors are accepted as model output. This includes removing non-numeric or stray characters, enforcing consistent spacing and bracket placement, and verifying that all indices lie within admissible bounds before integration into the closed-loop simulation.

Let $\{t_h\}_{h=0}^{\infty}$ denote the discrete decision epochs. At $h=0$, a one-time baseline assignment $\mathbf{Z}_0$ is computed using either the Hungarian algorithm or a randomized initialization. For all $h \ge 1$, the assignment is generated by the LLM-based reasoning policy. All scene variables are expressed in the common global Cartesian frame defined in Section~\ref{sec:problem}. The function \texttt{Assignment()} receives $\mathbf{S}_h$ and $\mathbf{Z}_{h-1}$ as inputs, builds the structured prompt $\mathcal{P}_h$ internally via \texttt{formatPrompt()}, sends it to \texttt{gpt-4o-mini}, and parses the first-line MATLAB vector response using RegEx before performing feasibility checks against~\eqref{eq:assign_constraints}.

\begin{algorithm}
\caption{High-level mission loop (baseline initialization at $h{=}0$)}
\label{alg:mission_loop}
\begin{algorithmic}[1]
\State Initialize interceptor, target, and asset states
\State Gather global scene data $\mathbf{S}_0$
\State $\mathbf{Z}_0 \gets$ \textbf{BaselineInit}$(\mathbf{S}_0)$ \hfill \% one-time Hungarian method
\For{$h = 1,2,\dots$}
    \State Gather global scene data $\mathbf{S}_h$ (position, velocity, and asset information)
    \State $\mathbf{Z}_h \gets \texttt{Assignment}(\mathbf{S}_h, \mathbf{Z}_{h-1})$ \hfill \% LLM-based WTA using $\mathcal{P}_h$
    \State Apply guidance law (e.g.,~\eqref{eq:png}) for each assigned interceptor–target pair during $t \in [t_h, t_{h+1})$
    \State Update all states and proceed to the next decision epoch
\EndFor
\end{algorithmic}
\end{algorithm}

The detailed LLM-driven decision process is outlined in \textbf{Algorithm~\ref{alg:llm_loop}}, which explicitly constructs the prompt, sends it to the model, parses the raw response, and outputs a validated assignment vector $\mathbf{Z}_h$.

\begin{algorithm}
\caption{Assignment($\mathbf{S}_h$, $\mathbf{Z}_{h-1}$)}
\label{alg:llm_loop}
\begin{algorithmic}[1]
\State Compute pairwise distances $D_{ih}(t_h)$ and relative closing speeds $C_{ih}(t_h)$
\State Estimate time-to-asset $\tau_h$ for each target
\State Collect scenario data $\theta_h$, $\psi_h$
\State Construct scene vector $\mathbf{S}_h = (D_{ih},C_{ih}, \tau_h, \theta_h, \psi_h, \mathbf{Z}_{h-1})$
\State Build prompt $\mathcal{P}_h = \mathrm{formatPrompt}(\mathbf{S}_h, \mathbf{Z}_{h-1})$
\State Send $\mathcal{P}_h$ to LLM; receive raw response $\mathcal{R}_h$
\State Parse $\mathcal{R}_h$ using RegEx; extract assignment vector $\mathbf{Z}_h$
\State Validate $\mathbf{Z}_h$ against~\eqref{eq:assign_constraints}
\State \Return $\mathbf{Z}_h$
\end{algorithmic}
\end{algorithm}

\begin{figure*}[t]
\centering
\begin{minipage}{0.95\textwidth}
\begin{llmprompt}{Example for prompt $\mathcal{P}_h$}
{\color{promptgreen}
\query{You are an expert mission planner for a weapon target assignment problem.\\
\textbf{Goal}: Solve the optimal assignment problem and protect high-priority assets by assigning interceptors to incoming targets.\\[3pt]
\textbf{PROVIDED DATA STRUCTURE:}\\
$N_i$ = number of interceptors (agents), $N_t$ = number of targets, $N_a$ = number of defended assets.\\
Agents: agent$_i$, $i=1,\dots,N_i$; Targets: target$_k$, $k=1,\dots,N_t$; Assets: asset$_m$, $m=1,\dots,N_a$.\\
\texttt{PREVIOUS\_ASSIGNMENT}: MATLAB row vector where entry $i$ gives the Target~ID assigned to Agent~$i$.\\
\texttt{DISTANCE\_MATRIX} $(N_i{\times}N_t)$: distance between Agent~$i$ and Target~$k$.\\
\texttt{CLOSING\_MATRIX} $(N_i{\times}N_t)$: relative closing speed between Agent~$i$ and Target~$k$.\\
\texttt{TIME\_TO\_ASSET} $(N_t)$: time until each target reaches its associated asset.\\
\texttt{THREAT\_LEVEL} $(N_t)$: threat level of each target.\\
\texttt{ASSET\_PRIORITY} $(N_a)$: priority of each defended asset.\\[3pt]
\textbf{CONSTRAINTS:}\\
- Each interceptor must be assigned to exactly ONE target.\\
- Returned vector must follow the same format as \texttt{PREVIOUS\_ASSIGNMENT}\\
  (index $i$ = Agent~ID, value = Target~ID).\\
- Avoid frequent reassignments; keep \texttt{PREVIOUS\_ASSIGNMENT} unless clearly advantageous.\\
- Prefer small distance, high closing speed, and low time-to-asset.\\
- Prioritize high-priority assets.\\
- RETURN ONLY a MATLAB row vector in the same format as \texttt{PREVIOUS\_ASSIGNMENT}.\\[3pt]
\textbf{CURRENT SCENARIO INFORMATION:}\\
$N_i = 10,\ N_t = 10,\ N_a = 3$\\
\texttt{PREVIOUS\_ASSIGNMENT:} [2 1 3 10 8 4 7 5 9 6]\\
\texttt{DISTANCE\_MATRIX:} [[4.1,4.5,\dots];\dots]\\
\texttt{CLOSING\_MATRIX:} [[1.3,0.8,\dots];\dots]\\
\texttt{TIME\_TO\_ASSET:} [9.2,7.8,\dots]\\
\texttt{THREAT\_LEVEL:} [0.9,0.6,\dots]\\
\texttt{ASSET\_PRIORITY:} [0.9,0.6,0.4]\\[3pt]
\textbf{DECISION REQUEST:}\\
Please return your decision for the assignment as a MATLAB row vector in the same format as \texttt{PREVIOUS\_ASSIGNMENT}, where index $i$ corresponds to the Agent~ID and the value corresponds to the assigned Target~ID. Example: [2 1 3 10 8 4 7 5 9 6].
}}
\end{llmprompt}
\end{minipage}
\caption{Example for structured prompt $\mathcal{P}_h$ used for reasoning-based assignment generation.}
\label{fig:prompt_structure}
\end{figure*}

\begin{figure}[t]
\centering
\begin{llmprompt}{Example \texttt{ChatGPT} response}
\query{
System: Please assign interceptors to the nearest targets.\\
}
\resp{
[2, 1, 3, 10, 8, 4, 7, 5, 9, 6]\\
``Reassigned I4 and I8 to reduce time-to-asset for the highest-priority threat while minimizing switches.''
}
\end{llmprompt}
\caption{Example for potential LLM response with a valid MATLAB-style assignment vector and a short reasoning explanation.}
\label{fig:placeholder}
\end{figure}

Figures~\ref{fig:prompt_structure} and~\ref{fig:placeholder} illustrate the structured prompt $\mathcal{P}_h$ and a representative model response, respectively. 
The prompt (Fig.~\ref{fig:prompt_structure}) contains three main components: a system-level description, current scene information, and an explicit decision request formatted according to MATLAB I/O conventions.

\begin{itemize}
    \item \textbf{System prompt:} Defines the problem structure and assignment constraints, ensuring that each interceptor is matched to exactly one target while minimizing unnecessary switches.
    \item \textbf{Scene prompt:} Encodes the scenario-dependent data such as distances, closing speeds, time-to-asset, threat levels, and priorities, along with the previous assignment vector.
    \item \textbf{Decision request:} Instructs the model to output only a MATLAB-style row vector that corresponds to the updated assignment, maintaining one-to-one indexing between agent and target IDs.
\end{itemize}

This format ensures deterministic model behavior and parseable responses. 
The MATLAB routine validates that the first line of the response is a syntactically correct row vector, removing non-numeric characters, verifying bracket and comma structure, and clipping indices to the admissible target range. 
Malformed or ambiguous responses trigger an automatic re-query, while repeated errors or timeouts lead to a fallback mechanism that computes $\mathbf{Z}_h$ using either the Hungarian or MILP solver. 
The interface operates with typical round-trip times between $1$–$2$\,s, maintaining real-time feasibility.

Between $t_h$ and $t_{h+1}$, the assignments remain constant and guidance follows PNG~\eqref{eq:png}. 
Unlike the classical surrogate-based formulation, the LLM interprets the full contextual information $(D_{ih},C_{ih},\tau_h,\theta_h,\psi_h,\mathbf{Z}_{h-1})$ directly, thereby eliminating manual weight calibration and capturing interdependent mission logic such as switching aversion and asymmetric urgency across assets. 
Empirical evaluations (Section~\ref{sec:results}) confirm that the resulting assignments are feasible, interpretable, and prioritize imminent high-value threats at real-time latencies.

\section{Simulation Results}
\label{sec:results}
For validation and proof of concept, we consider a two-dimensional defended-asset scenario with three protected assets, ten incoming red targets, and ten blue interceptors. The assets, representing stationary high-value locations such as air bases or command sites, are positioned close to the origin and surrounded by circular protection zones indicated by black dashed circles. Both interceptors and targets are modeled as double-integrator systems, and all states are expressed in a global Cartesian frame (in kilometers) centered around the defended area.

\begin{figure}
    \centering
    \includegraphics[width=\columnwidth]{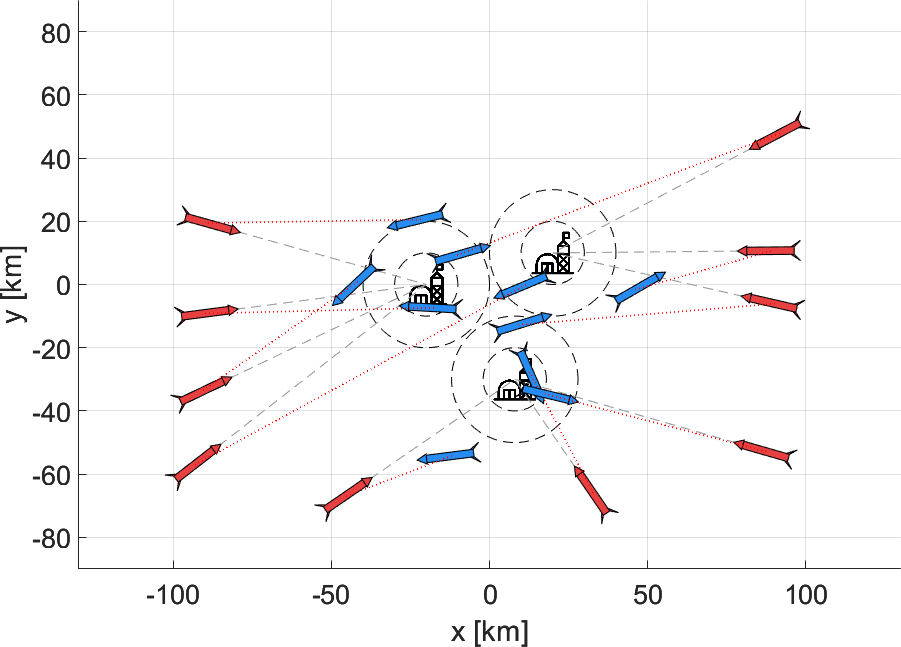}
    \caption{Initial scenario configuration ($t = 0\,\mathrm{s}$).}
    \label{fig:t0}
\end{figure}

At the beginning of the simulation (Figure~\ref{fig:t0}), all interceptors are assumed to be fully deployed and ready for engagement, with no launch delay or boost phase. The red arrows denote incoming targets, each moving toward one of the three protected assets along gray dashed lines that represent their intended impact trajectories. The blue arrows denote interceptors governed by the proposed LLM-based WTA framework. Red dotted lines indicate the dynamically assigned interceptor–target pairs at the initial time $t = 0\,\mathrm{s}$. Already at this stage, the model produces an assignment that closely aligns with an optimal geometric configuration, demonstrating coherent spatial reasoning.

\begin{figure}
    \centering
    \includegraphics[width=\columnwidth]{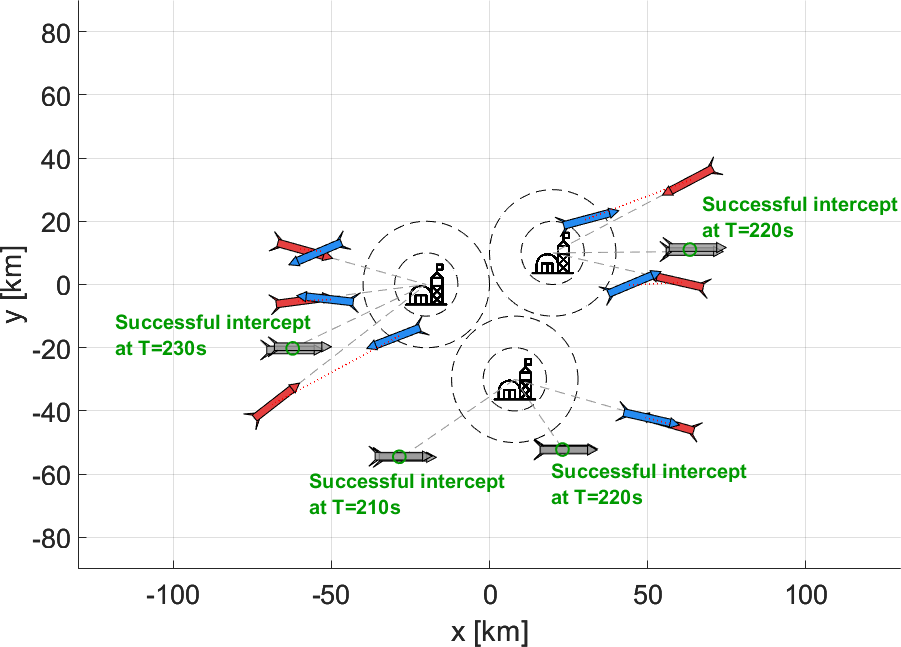}
    \caption{Intermediate mission stage ($t = 220\,\mathrm{s}$).}
    \label{fig:t1}
\end{figure}

As the mission progresses to approximately $t = 220\,\mathrm{s}$ (Figure~\ref{fig:t1}), several interceptors have already engaged their assigned targets using PNG~\eqref{eq:png}. The figure shows ongoing and completed engagements, with successful interceptions marked by small green circles labeled with the corresponding time of intercept (e.g., ``Successful intercept at $T = 220\,\mathrm{s}$''). Despite partial target elimination, the LLM maintains a consistent assignment for the remaining interceptors without unnecessary switching. This indicates that the model effectively interprets the dynamic scene and adapts the assignment policy as targets are removed from the engagement space.

\begin{figure}
    \centering
    \includegraphics[width=\columnwidth]{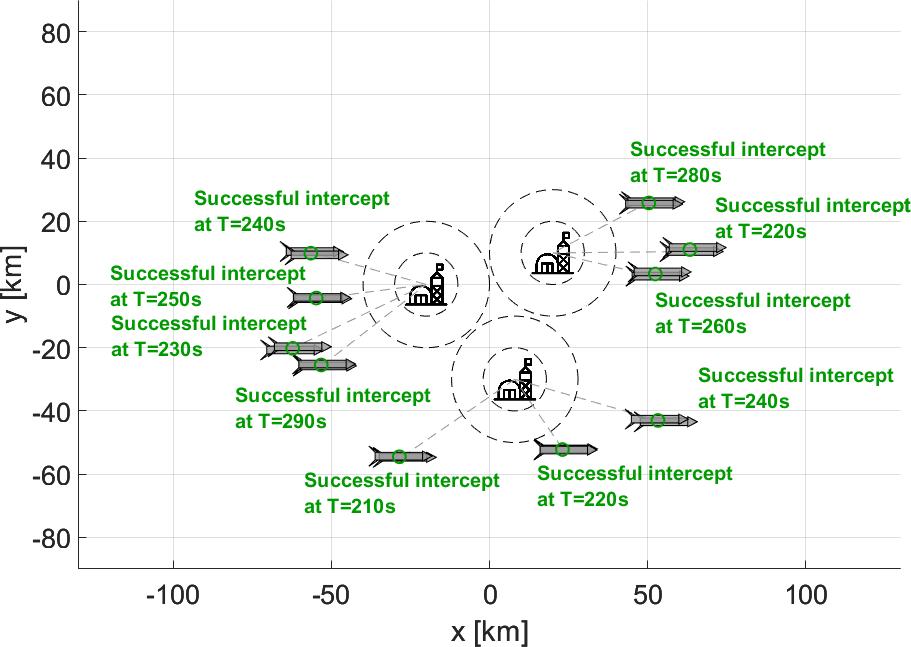}
    \caption{Final mission state ($t = 290\,\mathrm{s}$).}
    \label{fig:t2}
\end{figure}

By the final snapshot at $t = 290\,\mathrm{s}$ (Figure~\ref{fig:t2}), all targets have been successfully intercepted before reaching any of the protected zones. The system remains stable throughout the mission, with each blue interceptor completing its assigned engagement. The LLM maintains the same assignment consistency across decision epochs, showing that even without explicit optimization, reasoning-based policies can provide robust and interpretable mission-level behavior.

The results confirm that the proposed reasoning-based framework achieves consistent allocations, prevents unnecessary reassignment, and successfully protects all assets under uncertainty. Although the reasoning latency per LLM query ranges between $1-2\,\mathrm{s}$, the approach remains applicable for near-real-time decision support. In dynamic engagement settings, assignment loops typically operate at low update frequencies (approximately 0.5--1\,Hz), which makes the proposed reasoning-based architecture compatible with current control cycle requirements. Latency can be further reduced by employing more compact or locally hosted models fine-tuned for tactical reasoning.

For the considered scenario, the classical algorithms introduced in Section~\ref{sec:wta}, namely the Hungarian algorithm, MILP, and auction-based approaches, exhibited similar or comparable mission-level performance, confirming the validity of the LLM-based reasoning method relative to established optimization baselines. This demonstrates that, for the evaluated case, LLM-based reasoning can achieve decision quality equivalent to traditional optimization while providing improved interpretability and adaptability.

Importantly, all outcomes were achieved without any predefined weighting in the assignment cost, demonstrating that the LLM can exploit general reasoning capabilities to identify spatially and temporally coherent assignments. While the current setup involves non-maneuvering targets, future studies will extend the framework to cooperative and evasive target behaviors to assess adaptability under more complex adversarial conditions.

\section{Conclusion}
This paper introduced a reasoning-based framework for dynamic weapon–target assignment that integrates LLMs into the control-theoretic decision process of cooperative missile guidance. The approach replaces manually tuned surrogate cost functions with LLM-driven reasoning, allowing direct interpretation of mission-relevant information such as geometry, timing, and asset priority. By maintaining the classical decomposition between assignment and guidance, the framework remains compatible with existing control architectures while enhancing contextual adaptability.

Simulation studies in multi-asset defense scenarios demonstrated that the reasoning-based system provides consistent and interpretable allocations, reduces unnecessary reassignments, and preserves mission-level coherence under uncertainty. The integration of an LLM as a reasoning module enables flexible, human-like prioritization and robust decision-making across varying tactical conditions.

The presented results establish a proof of concept that high-level reasoning can complement or partially replace explicit optimization in control-oriented assignment problems. Future work will focus on thorough benchmarking against established optimization algorithms, assessing sensitivity to different LLM architectures, and developing hybrid frameworks that combine deterministic optimization with contextual reasoning to achieve reliable and explainable decision support in autonomous defense systems

\textbf{Acknowledgements.} The author would like to thank Carsten Schwarz from the DLR- Institute of Flight Systems, Department of Flight Dynamics and Simulation, for stimulating and insightful discussions on weapon–target assignment, which greatly contributed to the development of this work.

\bibliographystyle{IEEEtran}
\bibliography{./references}

\end{document}